\documentstyle[preprint,aps,epsfig]{revtex}
\draft
\begin{document}
\preprint{\it Nucl. Phys. A (2004) in press.}
\title{Effects of momentum-dependent symmetry potential on 
heavy-ion collisions induced by neutron-rich nuclei}
\bigskip

\author{\bf Bao-An Li$^{\rm a}$, 
Champak B. Das$^{\rm b}$\footnote{Present address: Variable Energy Cyclotron Center, 1/AF, Bidhannagar, Kolkata-700 064, India} 
Subal Das Gupta$^{\rm b}$ and Charles Gale$^{\rm b}$}
\address{$^{a}$Department of Chemistry and Physics\\
P.O. Box 419, Arkansas State University\\
State University, Arkansas 72467-0419, USA}

\address{$^{b}$Physics Department, McGill University, Montr{\'e}al, Canada H3A 2T8}
\date{\today}
\maketitle

\begin{abstract}
Using an isospin- and momentum-dependent transport model we study 
effects of the momentum-dependent symmetry potential on 
heavy-ion collisions induced by neutron-rich nuclei. 
It is found that symmetry potentials 
with and without the momentum-dependence but corresponding to the same density-dependent 
symmetry energy $E_{sym}(\rho)$ lead to 
significantly different predictions on several $E_{sym}(\rho)$-sensitive experimental 
observables especially for energetic nucleons. The momentum- and density-dependence of 
the symmetry potential have to be determined simultaneously in order to extract the 
$E_{sym}(\rho)$ accurately. The isospin asymmetry of midrapidity nucleons at high transverse 
momenta is particularly sensitive to the momentum-dependence of the symmetry potential. 
It is thus very useful for investigating accurately the 
equation of state of dense neutron-rich matter.
\end{abstract}

\pacs{25.70.-z, 25.70.Pq., 24.10.Lx}
\maketitle

\section{Introduction}
The equation of state (${\rm EOS}$) of asymmetric nuclear matter at density $\rho$ and isospin asymmetry
$\delta\equiv(\rho_{n}-\rho _{p})/(\rho _{p}+\rho _{n})$ is usually expressed as
\begin{equation}\label{ieos}
E(\rho ,\delta )=E(\rho ,\delta =0)+E_{\text{\textrm{sym}}}(\rho )\delta
^{2}+{\cal O}(\delta^4),
\end{equation}
where $E_{\rm sym}(\rho)$ is the nuclear symmetry energy. 
The latter is very important for many interesting astrophysical problems and yet 
poorly known for dense neutron-rich matter\cite{bethe,lat01,ibook01}. Predictions based on various 
many-body theories diverge widely, especially at high densities. In fact, even the sign of 
the symmetry energy above $3\rho_0$ is still very uncertain\cite{bom1}. 
Since nuclear reactions induced by high energy radioactive beams can produce transiently
dense neutron-rich matter, the planned Rare Isotope Accelerator (RIA) and the 
new accelerator facility at GSI provide the first opportunity in terrestrial laboratories to explore 
experimentally the ${\rm EOS}$ of dense neutron-rich 
matter\cite{ireview98,dan02,li03,nsac02}. Crucial to the extraction of critical information 
about the $E_{\rm sym}(\rho)$ is the comparison between experimental data and transport model 
calculations. Based on isospin-dependent transport model calculations, several experimental 
observables have been identified as promising probes of the $E_{sym}(\rho)$, 
such as, the neutron/proton ratio \cite{li97}, isoscaling in nuclar 
multifragmentation\cite{betty,tan01,bar02,betty03}, the neutron-proton 
differential flow \cite{li00,gre03,sca}, the neutron-proton correlation function\cite{chen03}
and the isobaric yield ratios of light clusters\cite{chen03b}. 

An important input to the transport models are the isovector (symmetry) and isoscalar potentials. 
However, in all transport models for heavy-ion collisions the momentum-dependence of 
the symmetry potential was seldom taken into account. In the recent work of ref.\cite{riz}
effects of the neutron-proton effective mass splitting due to the momentum-dependent nuclear potentials 
were studied using Bombaci's parameterization of single nucleon potential\cite{bom1}. 
Nevertheless, effects of the momentum-dependence of the symmetry potential 
in heavy-ion collisions are still largely unknown. On the other hand, effects of the 
momentum-dependence of the isoscalar potential in nuclear reactions are 
well-known\cite{gbd87,pra88,gale90,bla93,pawel91,fai,zhang94}. 
This is mainly because only very recently was the momentum-dependence of the symmetry potential 
given in a form practically usable in transport model calculations\cite{das03}. 
In this work, we study in detail effects of the momentum-dependence of the symmetry potential 
on heavy-ion collisions induced by neutron-rich nuclei. A short report of this study can be found 
in ref.\cite{lidas03}. In the next section, we discuss the momentum-dependence of the symmetry potential
using as a guide results of Hartree-Fock calculations with Gogny effective interactions. The strengths 
of the symmetry potential with and without the momentum-dependence are compared. 
In section III, we present and discuss effects of the momentum-dependence of the symmetry potential
on several experimental observables in heavy-ion collisions induced by neutron-rich nuclei. 
A summary is given at the end.

\section{Momentum-dependence of the symmetry potential}
That the momentum-dependence of the single particle potential is different for neutrons and protons 
in asymmetric nuclear matter is well-known and has been a subject of intensive research based on various 
many-body theories using non-local interactions, see e.g., ref. \cite{bom1} for a review. 
Experimentally, it is known that the strength of nuclear symmetry potential 
decreases linearly from about $28\pm 6$ MeV with increasing 
kinetic energy with a slope of about 0.1-0.2 from optical model analyses of nucleon-nucleus 
scattering data\cite{hod94}. However, the available data is limited only to 
nucleon energies less than 100 MeV. Guided by a Hartree-Fock calculation using the 
Gogny effective interaction, the single nucleon 
potential was recently parameterized as \cite{das03} 
\begin{eqnarray}\label{mdi}
U(\rho,\delta,\vec p,\tau) &=& A_u\frac{\rho_{\tau'}}{\rho_0}
+A_l\frac{\rho_{\tau}}{\rho_0}\nonumber \\ &+&
B(\frac{\rho}{\rho_0})^{\sigma}(1-x\delta^2)-8\tau x\frac{B}{\sigma+1}\frac{\rho^{
\sigma-1}}{\rho_0^{\sigma}}\delta\rho_{\tau'} \nonumber \\
&+&\frac{2C_{\tau,\tau}}{\rho_0}
\int d^3p'\frac{f_{\tau}(\vec r,\vec p')}{1+(\vec p-\vec p')^2/\Lambda^2}\nonumber \\
&+&\frac{2C_{\tau,\tau'}}{\rho_0}
\int d^3p'\frac{f_{\tau'}(\vec r,\vec p')}{1+(\vec p-\vec p')^2/\Lambda^2}
\end{eqnarray}

The parameter $x$ can be adjusted to mimic predictions on the density dependence of the 
symmetry energy $E_{sym}(\rho)$ by microscopic and/or phenomenlogical many-body theories.
In the above $\tau=1/2$ ($-1/2$) for neutrons (protons) and $\tau\neq\tau'$; $\sigma=4/3$;
$f_{\tau}(\vec r,\vec p)$ is the phase space distribution 
function at coordinate $\vec{r}$ and momentum $\vec{p}$. 
The parameters $A_u, A_l, B, C_{\tau,\tau}, C_{\tau,\tau'}$ and $\Lambda$ 
were obtained by fitting the momentum-dependence of the 
$U(\rho,\delta,\vec p,\tau)$ predicted by the Gogny Hartree-Fock and/or the Brueckner-Hartree-Fock 
calculations, saturation properties of symmetric nuclear matter and the symmetry energy of 30 MeV 
at normal nuclear matter density $\rho_0=0.16/fm^3$\cite{das03}. 
More specifically, B=106.35 MeV and $\Lambda=p_F^0$ is the nucleon Fermi momentum in symmetric 
nuclear matter.  $A_u$ and $A_l$ depend on the $x$ parameter according to
\begin{eqnarray}
A_u(x)&=&-95.98-x\frac{2B}{\sigma+1},\\
A_l(x)&=&-120.57+x\frac{2B}{\sigma+1}.
\end{eqnarray}
The compressibility of symmetric 
nuclear matter $K_{0}$ is set to be 211 MeV. 

The momentum-dependence of the symmetry potential 
steams from the different interaction strength parameters $C_{\tau,\tau'}$ and $C_{\tau,\tau}$ 
for a nucleon of isospin $\tau$ interacting, respectively, with unlike and like nucleons in 
the background fields. More specifically, we obtained $C_{unlike}=-103.4$ MeV and $C_{like}=-11.7$ MeV. 
One characteristic of the momentum dependence of the symmetry potential is the different effective
masses for neutrons and protons in isospin asymmetric nuclear matter. In terms of the single particle
potential $U(\rho,\delta,\vec p,\tau)$ the nucleon effective mass is defined as
\begin{equation}
\frac{m^*_{\tau}}{m_{\tau}}=\left[1+\frac{m_{\tau}}{p}\frac{dU}{dp}\right]^{-1}_{p_F^{\tau}}.
\end{equation}
Since the momentum dependent part of the single particle potential in eq. \ref{mdi} 
is independent of the 
parameter $x$, the nucleon effective mass is the same for different symmetry energies. 
Shown in Fig.\ 1 are the nucleon effective masses as a function of density (upper window) 
and isospin asymmetry (lower window). These results are consistent with the Bruckner Hartree-Fock
predictions\cite{bom91}. 
\begin{figure}[ht]
\centering \epsfig{file=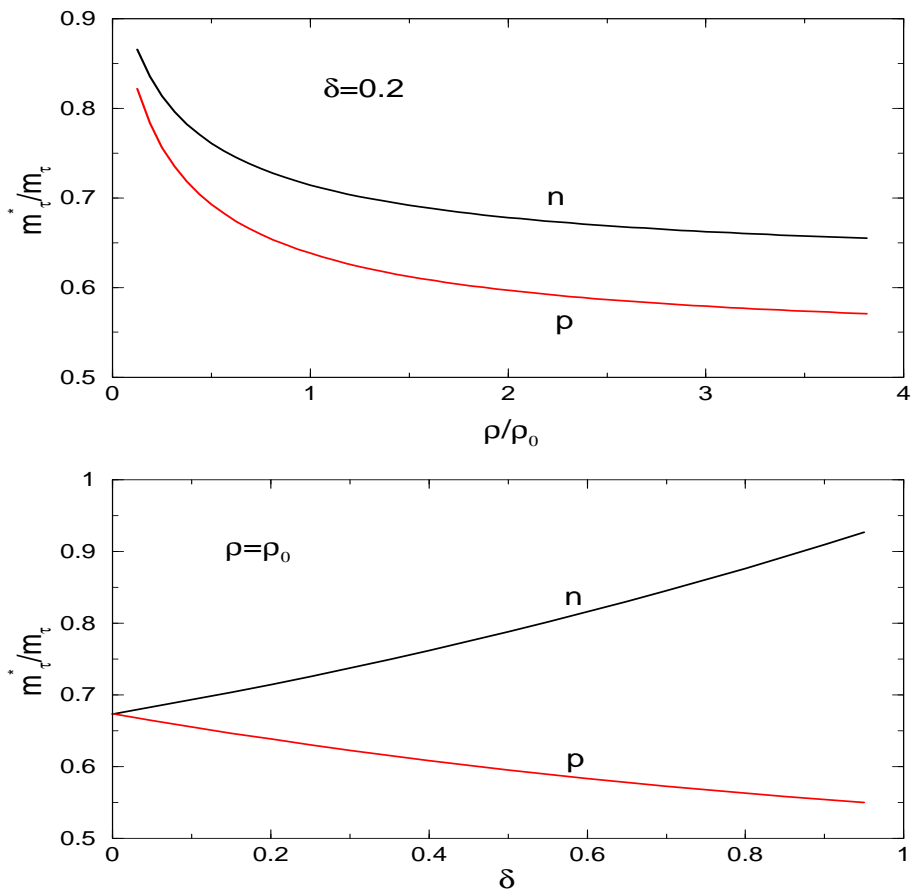,width=12cm,height=12cm,angle=-90} 
\vspace{1.cm}
\caption{{\protect\small Nucleon effective masses as a function of density (upper window) and
isospin asymmetry (lower window).}}
\label{esym}
\end{figure}

The parameter $x$ was introduced to reflect the largely uncertain behavior of $E_{sym}(\rho)$
at supranormal densities, specifically its potential part $E_{sym}^p(\rho)$.  
The latter corresponding to $x=1$ (denoted by MDI(1) which gives the same $E_{sym}(\rho)$
as the default Gogny interaction) and 
$x=0$ (MDI(0)) are shown in Fig.\ 2. For comparisons we have also included the potential 
part of the more stiffer symmetry energy $E_{sym}^{RMF}(\rho)\equiv30\rho/\rho_0$, 
predicted by several models, 
such as, the Relativistic-Mean-Field (RMF) model. As a reference, the kinetic contribution 
$E_{sym}^k(\rho)=\hbar^2/6m_n(3\pi^2\rho/2)^{2/3}$ is also shown. 
The variation of $E_{sym}(\rho)$ resulted by changing from the MDI(1) to the MDI(0) parameter set
is well within the uncertain range predicted by various many-body theories, 
see e.g., refs.\cite{brown,vmb,bom91}. We found that the potential contribution to 
the symmetry energy can be parameterized as 
\begin{equation}\label{mdi1}
E_{sym}^p(\rho)=3.08+39.6u-29.2u^2+5.68u^3-0.523u^4
\end{equation}   
for the MDI(1) (Gogny) interaction and
\begin{equation}\label{mdi0}
E_{sym}^p(\rho)=1.27+25.4u-9.31u^2+2.17u^3-0.21u^4
\end{equation}   
for the MDI(0) interaction, where $u\equiv \rho/\rho_0$ is the reduced nucleon density.
\begin{figure}[ht]
\centering \epsfig{file=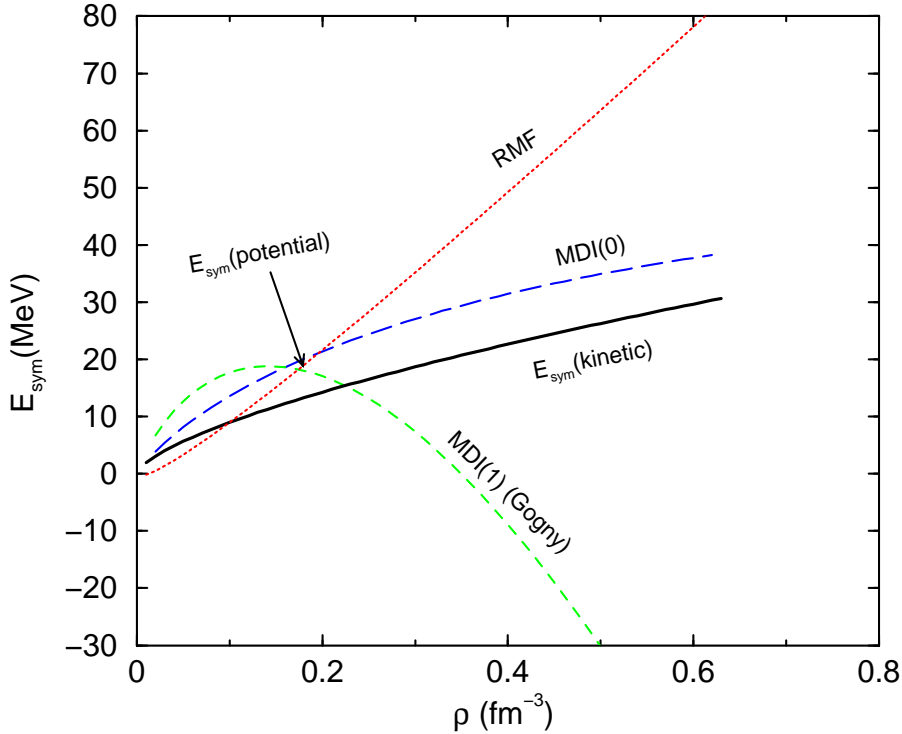,width=10cm,height=12cm,angle=-90} 
\vspace{1.cm}
\caption{{\protect\small The density dependence of the 
potential and kinetic parts of nuclear symmetry energy.}}
\label{esym}
\end{figure}

To compare with the experimentally found momentum-dependence of the symmetry 
potential in nucleon-nucleus scatterings, we illustrate in Fig.\ 3 the strength 
of the symmetry potential $(U_n-U_p)/2\delta$ as a function of nucleon kinetic energy at densities of 
$0.5\rho_0$, $\rho_0$ and $2\rho_0$, for $x=1$ (left) and $x=0$ (right), respectively.
The results are obtained by using $\delta=0.2$. However, since the isovector part of the nucleon 
potential is approximately propotional to $\delta$, the quantity $(U_n-U_p)/2\delta$ is 
approximately independent of $\delta$. At normal density, for both $x=1$ and $x=0$ the 
strength of symmetry potential decreases from about 33 MeV at $E_{kin}=0$ with a slope of about 0.22 for 
$E_{kin}\leq 100$ MeV in general agreement with the nucleon-nucleus scattering data. Moreover, 
it is seen that the strength approaches zero above about 200 MeV. At other densities, 
however, the predictions are dramatically different for $x=1$ and $x=0$ especially 
at high densities. For $\rho=2\rho_0$, for instance, 
the strength of symmetry potential is negative and decreases quickly with energy 
indicating a stronger attractive/repulsive symmetry potentials for neutrons/protons with $x=1$. 
With $x=0$, however, at the same density $\rho=2\rho_0$ the strength of the symmetry potential is
now positive indicating the repulsive/attractive nature of the symmetry potential for neutrons/protons.
It is thus characteristically different from the $x=1$ case. It is also seen that the 
strength of the symmetry potential decreases relatively slowly with $x=0$. 
These different energy dependences of the symmetry potential 
are correlated to the negative and positive potential contribution to the 
symmetry energy at high densities for $x=1$ and $x=0$, respectively.
\begin{figure}[ht]
\centering \epsfig{file=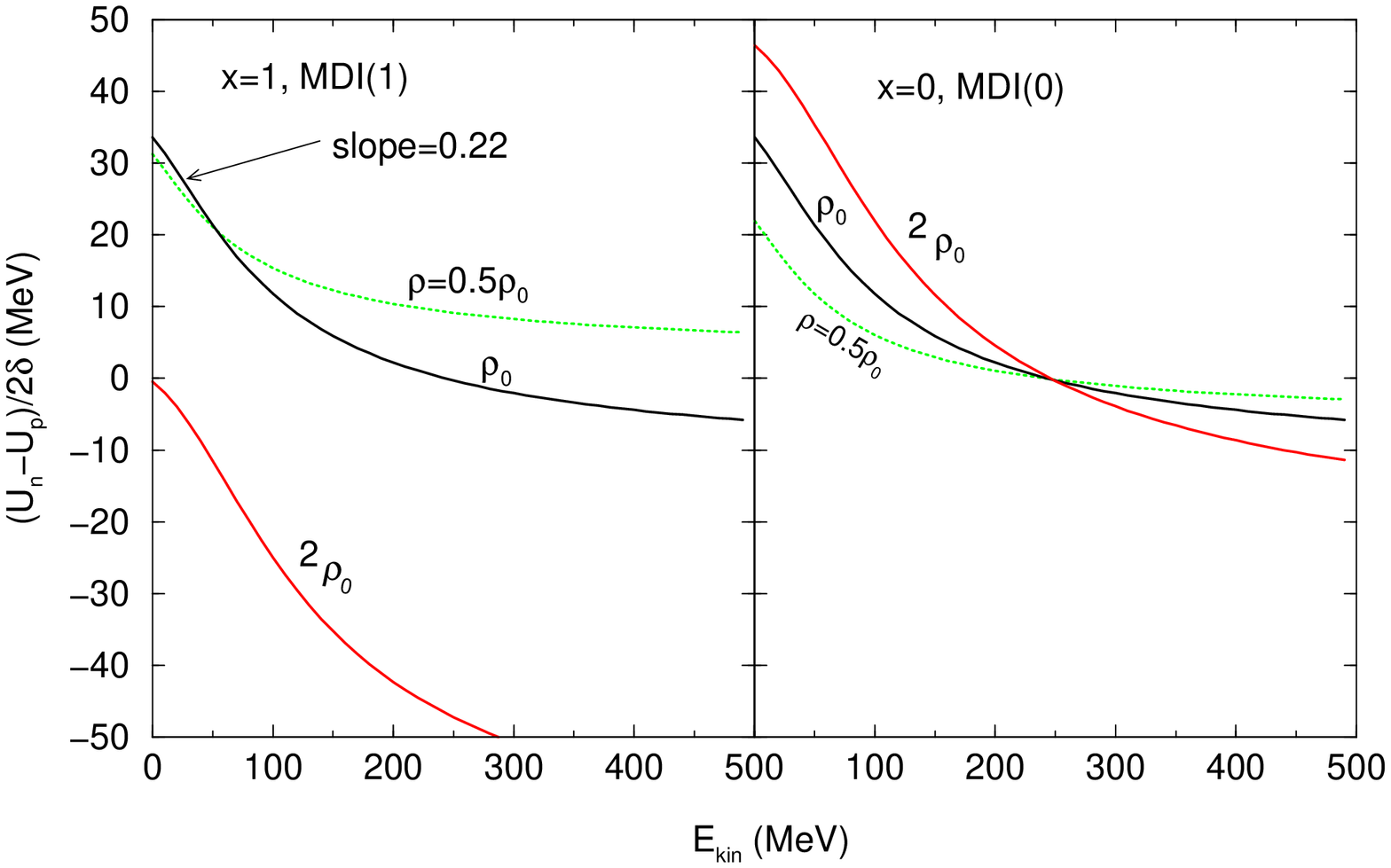,width=10cm,height=12cm,angle=-90} 
\vspace{1.cm}
\caption{{\protect\small Strength of the symmetry potential as a function of kinetic energy at three
densities for $x=1$ (left) and $x=0$ (right).}}
\end{figure}   
Since the C-terms in the single particle potential depend inseparably on the density,
momentum and isospin, to investigate effects of the momentum-dependence of the symmetry potential we shall
also compare results obtained using eq. \ref{mdi} with those obtained using a single nucleon potential 
$U_{noms}(\rho,\delta,\vec{p},\tau)\equiv U_0(\rho,\vec{p})+U_{sym}(\rho,\delta,\tau)$ that
has almost the same momentum-dependent isoscalar part $U_0(\rho,\vec{p})$ as that embedded 
in eq. \ref{mdi} and a momentum-${\it independent}$ symmetry potential $U_{sym}(\rho,\delta,\tau)$ 
that gives the same $E_{sym}(\rho)$ as eq.\ \ref{mdi}.
The momentum-independent $U_{sym}(\rho,\delta,\tau)$ can be obtained 
from 
\begin{equation}
U_{sym}(\rho,\delta,\tau)=\partial W_{sym}/\partial \rho_{\tau}, 
\end{equation}
where $W_{sym}$ is the isospin-dependent part of the potential energy density 
\begin{equation}
W_{sym}=E_{sym}^p(\rho)\cdot\rho\cdot\delta^2.
\end{equation}
For the symmetry energy $E_{sym}^{RMF}(\rho)$, 
for instance, one obtains 
\begin{equation}
U_{sym}^{RMF}(\rho,\delta,\tau)=4\tau\delta(30u-12.7u^{2/3})+4.23u^{2/3}\delta^2.
\end{equation} 
Using the parameterizations for $E_{sym}^p(\rho)$ in eqs. \ref{mdi1} and \ref{mdi0}, 
we obtain
\begin{eqnarray}
&&U^{MDI1}_{sym}(\rho,\delta,\tau)=4\tau\delta(3.08+39.6u-29.2u^2 +5.68u^3
\nonumber \\ && -0.52u^4)
-\delta^2(3.08+29.2u^2-11.4u^3+1.57u^4)
\end{eqnarray} 
for the MDI(1) parameter set and 
\begin{eqnarray}
&&U^{MDI0}_{sym}(\rho,\delta,\tau)=4\tau\delta(1.27+25.4u-9.31u^2 +2.17u^3
\nonumber \\ && -0.21u^4)
-\delta^2(1.27+9.31u^2-4.33u^3+0.63u^4)
\end{eqnarray}  
for the MDI(0) parameter set. The momentum-dependent isoscalar 
potential $U_0(\rho,\vec{p})$ will be given in the following shortly. 

The strengths of the symmetry potentials with and without the momentum-dependence 
but corresponding to the same $E_{sym}(\rho)$ are compared in Fig.\ 4 
by examining the difference between neutron and proton potentials $(U_n-U_p)/2\delta$. 
Again, this quantity is approximately independent of the parameter $\delta$.
The symmetry potential without the momentum-dependence is 
higher in magnitude and has generally steeper slopes than the 
momentum-dependent one for $\rho/\rho_0\leq 2.3$ 
with the MDI(1) parameter set and at all densities for the MDI(0) parameter set.
Moreover, the strength of momentum-dependent symmetry potentials decreases with 
the increasing momentum. Thus the difference between
the symmetry potentials with and without the momentum-dependence is larger for nucleons
with higher momenta.
\begin{figure}[ht]
\centering \epsfig{file=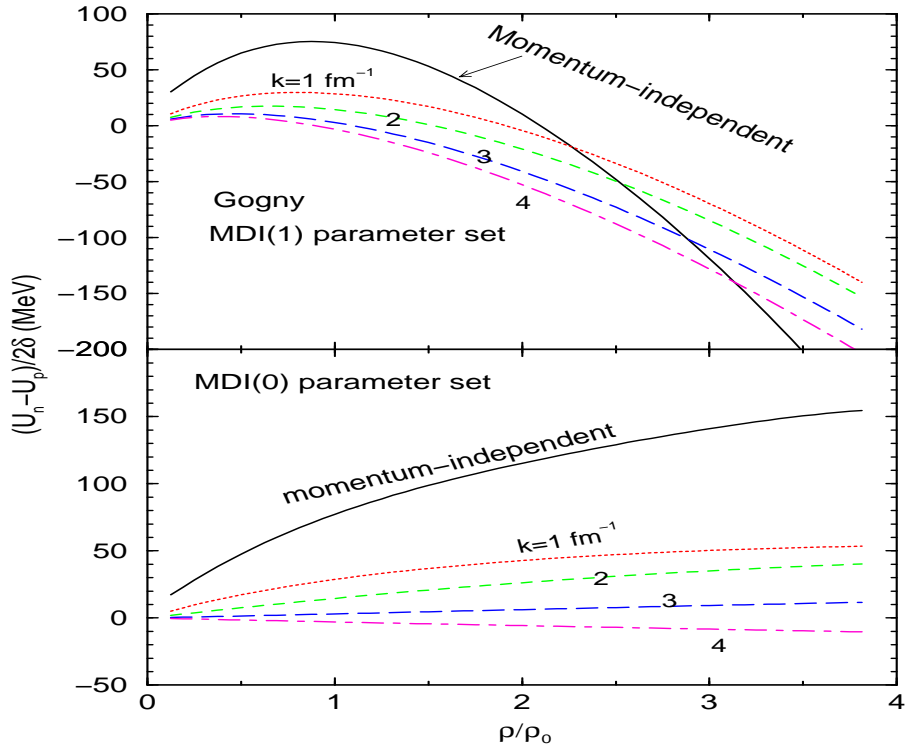,width=10cm,height=12cm,angle=-90} 
\vspace{1.cm}
\caption{{\protect\small.Strengths of the symmetry potentials with and without (solid) 
the momentum-dependence as a function of density as measured by the difference 
between neutron and proton potentials.}}
\label{usym}
\end{figure}  

To identify reliably effects of the momentum-dependence of the symmetry potential  
without much interference from density effects the $U_{noms}(\rho,\delta,\vec{p},\tau)$ 
should lead to about the same reaction dynamics 
and evolution of density profiles as the single particle potential in eq.\ \ref{mdi}.
Both of them are mainly determined by the isoscalar potential for which 
we select the original momentum-dependent Yukawa interaction (MDYI)\cite{gale90} 
\begin{eqnarray}\label{mdyi}
U_0(\rho,\vec{p})&=&-110.44u+140.9u^{1.24}\nonumber \\
&&-\frac{130}{\rho_0}
\int d^3p'\frac{f(\vec r,\vec p')}{1+(\vec p-\vec p')^2/(1.58p_F^0)^2},
\end{eqnarray}
where $p_F^0$ is the Fermi momentum. The compressibility $K_0$ for this interaction is 215 MeV.
This potential is compared with the variational calculation using the $UV14+UVII$ interactions in Fig.\ 5.
Except at high densities but low momenta the MDYI interaction gives about the same isoscalar potential
as the variational approach as first found in ref.\cite{gale90}. 
\begin{figure}[ht]
\centering \epsfig{file=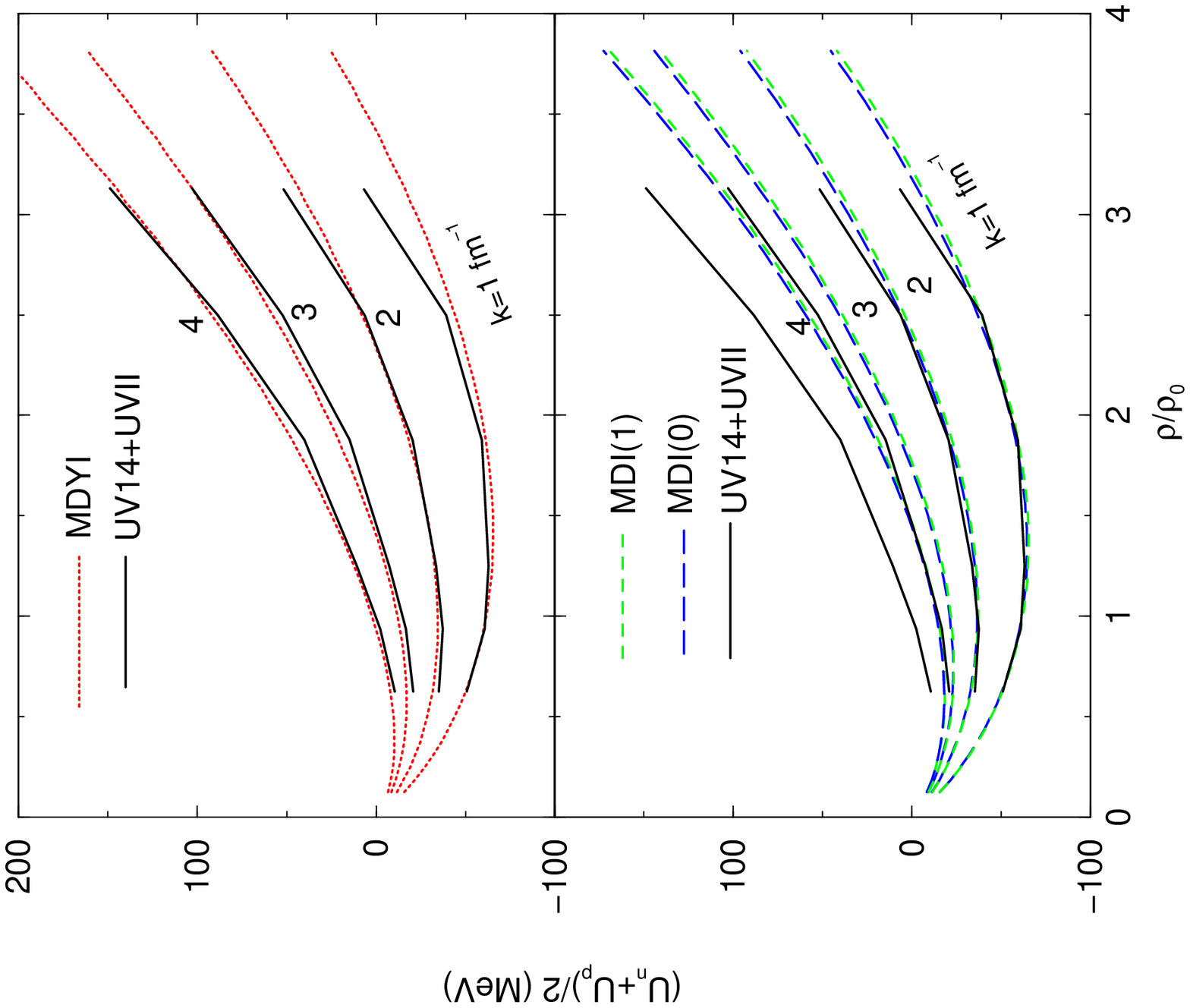,width=12cm,height=12cm,angle=-90} 
\vspace{1.cm}
\caption{{\protect\small Strengths of the isoscalar potential as 
a function of density at the four wave numbers for the MDI(1) and MDI(0) parameter 
sets (lower window) as well as the MDYI (upper window) interaction 
in comparison with the variational many-body calculations using the $UV14$ two-body
and the $UVII$ three-body potentials.
(upper window) }}
\end{figure}
In the lower window of Fig.\ 5 we compare the isoscalar part $(U_n+U_p)/2$ 
of the single nucleon potentials corresponding to the MDI(1) and MDI(0) parameter sets. 
It is seen that they have roughly the same values at all wave numbers $k$ examined, except at high 
densities where the MDI(0) parameter set leads to slightly stiffer isoscalar potentials. This feature 
ensures that the differences we see in reaction calculations using the MDI(1) and MDI(0) parameter
sets are truly due to the different symmetry potentials. As a reference we have also included Wiringa's
calculations of the isoscalar potentials within the variational many-body approach using the $UV14$ two-body
potential and the $UVII$ three-body potential. The single nucleon potential was obtained by using 
a realistic Hamiltonian that fits nucleon-nucleon scattering data, few-body nuclear binding 
energies and nuclear matter saturation properties\cite{wiringa88}. It was shown that combinations 
of different two-body and three-body forces lead to slightly different single nucleon 
potential especially at high densities and/or high momenta. It is seen that for wave numbers less 
than about $2fm^{-1}$ the MDI(1) and MDI(0) results are very close to that of the variational calculations.
At higher wave numbers, however, the variational results are higher. For our studies of isospin effects
in this work isoscalar potentials are less important as long as the reaction dynamics is 
about right. We thus consider the isoscalar parts of both the MDI(1) and MDI(0) interactions 
as very reasonable for the purpose of this work.  

Overall, the MDYI interaction leads to about the same isoscalar potential as the MDI(1) 
and MDI(0) interactions for $k$ less than about $2fm^{-1}$. We shall also verify numerically 
that the potential $U_{noms}(\rho,\delta,\vec{p},\tau)$ constructed this way indeed leads 
to about the same reaction dynamics and the evolution of density profiles 
as the potential in eq. \ref{mdi}.
For comparisons and identify effects due to the momentum-dependence and stiffness 
of the isoscalar potential, we use occasionally also the momentum-independent 
Bertsch-Kruse-DasGupta (BKD) isoscalar potentials.
The soft BKD is given by\cite{bkd} 
\begin{equation}
U_0(\rho)=-356u+303u^{7/6},
\end{equation}
and stiff BKD is given by
\begin{equation}
U_0(\rho)=-124u+70.5u^{2}.
\end{equation}
The compressibilities of these two potentials are $K_0=201$ and 377 MeV, respectively.

\section{Results and discussions}
In this section we examine effects of the momentum-dependence of the symmetry potential 
on the reaction dynamics and several experimental observables.  
Several observables are known to be sensitive 
only to the symmetry potential but not to the isoscalar potential. 
They are mainly neutron-proton differential or relative 
quantities\cite{li97,li00} where effects of the isoscalar potential 
with or without the momentum-dependence are largely canceled out.
Moreover, these observables are also not sensitive to the in-medium
nucleon-nucleon cross sections\cite{li97,chen03,chen03b} since the latter act
approximately identically on both neutrons and protons. 
Assuming the symmetry potential is momentum-independent, these observables 
were previously proposed as promising probes of the density-dependence of the symmetry energy. 
Our emphasis in this work will be on searching for experimental observables 
that can be used to determine simultaneously both the density- and momentum-dependence 
of the symmetry potential, and thus to determine uniquely 
the density-dependence of the symmetry energy. These studies are carried out within an 
isospin-dependent transport model\cite{li97}, for the latest review of this model we refer the 
reader to ref.\cite{li03}. In this work we concentrate on $^{132}$Sn+$^{124}$Sn reactions 
at the highest RIA beam energy of 400 MeV/nucleon. 

\subsection{Effects on reaction dynamics}
Shown in Fig.\ 6 are the evolutions of central baryon densities in $^{132}$Sn+$^{124}$Sn 
reactions at a beam energy of 400 MeV/nucleon and an impact parameter of 1 fm with 
various potentials. In the upper window, we first compare results obtained by using  
the soft BKD (solid thin line) and the MDYI (dot-dashed line) isoscalar potentials but  
the same symmetry potential $U_{sym}^{RMF}(\rho)$. It is seen that the soft BKD 
leads to a higher compression by about 10\% compared to the MDYI interaction in agreement
with previous findings in many studies. 
\begin{figure}[ht]
\centering \epsfig{file=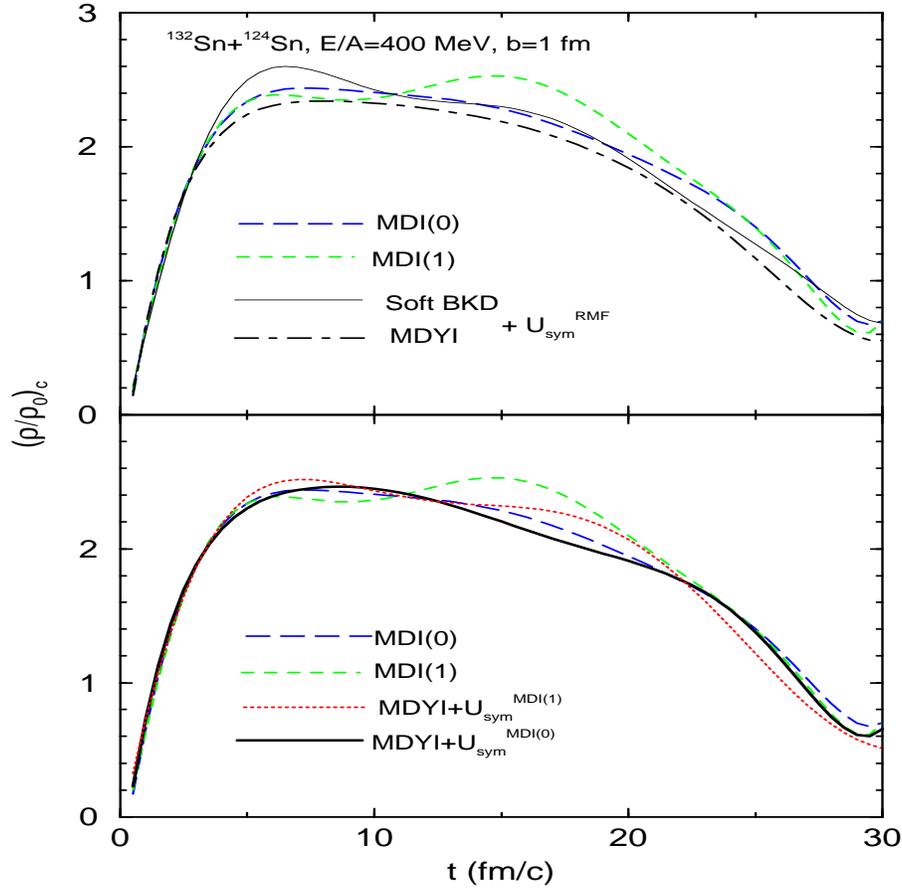,width=12cm,height=12cm,angle=-90} 
\vspace{1.cm}
\caption{{\protect\small Evolution of central baryon density with various potentials 
in the $^{132}Sn+^{124}Sn$ reaction at a beam energy of 400 MeV/nucleon and an 
impact parameter of 1 fm.}}
\label{figure5}
\end{figure}
Next, we compare results obtained with the MDI(0) (long dashed line) and 
the MDI(1) (dashed line) potentials. As we have discussed earlier they have about the same isoscalar 
potential while very different symmetry potentials as shown in Fig. 4. It is interesting to notice
that starting at about 10 fm/c when the central density is above about $2.4\rho_0$, the central density 
for the MDI(1) starts increasing significantly with respect to the MDI(0) case. 
This is what we have expected based on eq.\ 1 and the density dependence of the symmetry energy.
The relative values of the central densities reflect closely the relative stiffnesses of 
the underlying equations of state. The above observation is due to a softening of the underlying
EOS in the MDI(1) case. As shown in Fig.\ 2, the potential part of the symmetry energy increases 
continuously for the MDI(0) while decreases with increasing density above $\rho_0$ for the 
MDI(1) case. In fact, starting at about $2.4\rho_0$, the symmetry energy becomes negative 
with the MDI(1) parameter set, leading thus to the more compressed central region. Overall, however, the
differences in central densities using different potentials are less than about 10\%. 

In the lower window of Fig.\ 6, we study effects of the momentum-dependence of the symmetry potential on 
the central density. This is done by comparing results obtained by 
using the MDI(0) with MDYI+$U_{sym}^{MDI(0)}$, and MDI(1) with MDYI+$U_{sym}^{MDI(1)}$, respectively. 
The two pairs of potentials correspond to the same symmetry energy of MDI(0) and MDI(1), respectively.
It is seen that without the momentum-dependence of the symmetry potential, i.e., 
MDYI+$U_{sym}^{MDI(1)}$ and MDYI+$U_{sym}^{MDI(0)}$ cases, the central density is 
slightly lower. This is because the symmetry potential is effectively 
stronger without the momentum-dependence as shown in Fig.\ 4. For neutron-rich matter this effectively
stiffens the EOS, leading therefore to a lower compression. However, the general 
dynamics of the reaction is merely affected by the momentum-dependence of 
the symmetry potential. 

\subsection{Effects on nucleon emissions}
The generally repulsive/attractive symmetry potential for 
neutrons/protons at subnormal densities 
is to cause more neutrons/protons to be free/bound. 
We identify free nucleons as those with local densities less than $\rho_0/8$. 
At midrapidity, the isospin asymmetry of free nucleons are dominated by the high-density
behavior of nuclear symmetry energy. As the symmetry energy becomes more stiffer 
from the MDI(1) through MDI(0) to the $E_{sym}^{RMF}(\rho)$ as shown in Fig.\ 2, 
gradually more/less neutrons/protons are expected to be emitted at midrapidity. 
The different symmetry energies will thus lead to quite different isospin asymmetries 
of free nucleons at mid-rapidity. Shown in the upper window of Fig.\ 7 are the nucleon 
rapidity distributions with the four different potentials. Since the three potentials 
MDI(1), MDI(0) and $MDYI+U_{sym}^{RMF}(\rho)$ 
have about the same isoscalar part while their isovector 
parts are quite different, the observed difference among results using them 
clearly reveals the role of the symmetry potential in nucleon emissions. We also included the 
calculation with the $SBKD+U_{sym}^{RMF}(\rho)$ interaction as a reference.
First of all, it is seen that midrapidity nucleons are most sensitive to the variation of 
symmetry energy as one expects since these nucleons are the ones having 
gone through the high density phase. Moreover,
neutron rapidity distributions are more affected by the symmetry energy than protons. 
The dN/dy values around midrapidity are higher for neutrons with the stiffer symmetry 
energies as one expects. 
\begin{figure}[ht]
\centering \epsfig{file=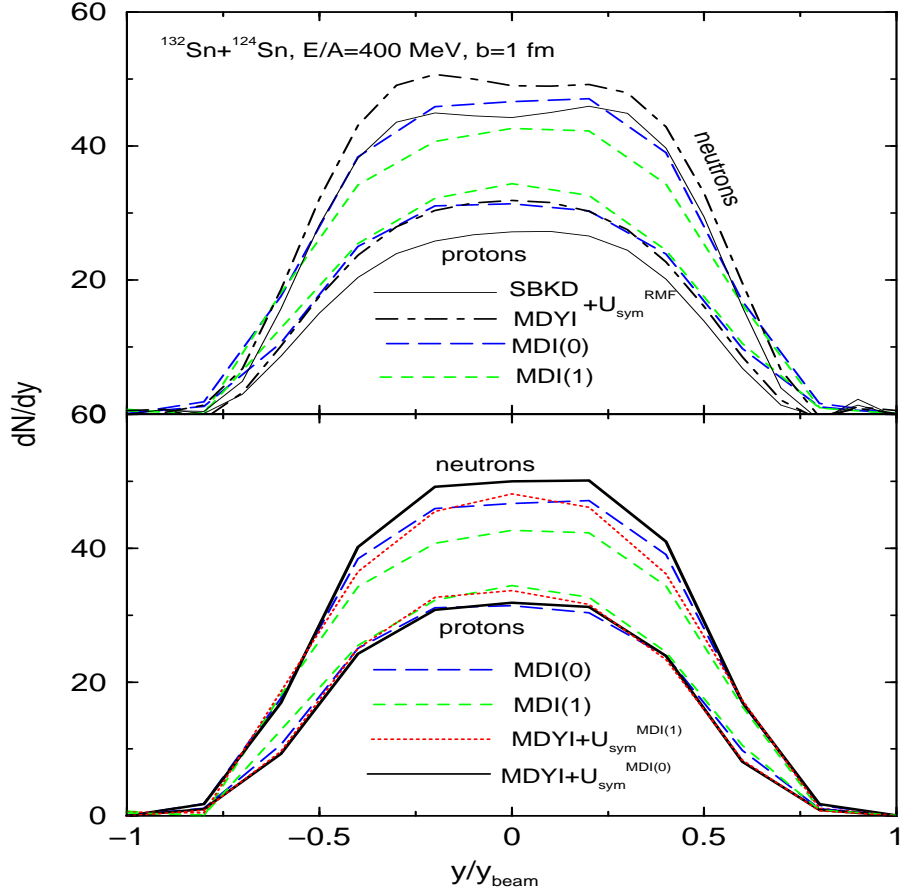,width=12cm,height=12cm,angle=-90} 
\vspace{1.cm}
\caption{{\protect\small Rapidity distributions of neutrons and protons  
with various potentials in the $^{132}Sn+^{124}Sn$ reaction at a beam energy of 
400 MeV/nucleon and an impact parameter of 1 fm.}}
\label{figure6}
\end{figure}
Included in the upper window Fig.\ 7 are also results obtained with 
the soft BKD isoscalar potential and the $U_{sym}^{RMF}(\rho)$ symmetry potential. 
Compared to the calculations with the MDYI isoscalar 
potential but the same symmetry potential $U_{sym}^{RMF}(\rho)$, the soft BKD leads to reduced 
emissions of both neutrons and protons. In particular, the resulting neutron rapidity distribution is
very close to the MDI(0) case. This illustrates one serious drawback of inclusive observabes
for studying the isospin-dependence of the nuclear EOS, such as the nucleon rapidity distributions 
studied here, that they are sensitive to both the isoscalar and isovector potentials. 
This is an important reason for us to concentrate on observables that are only sensitive 
to the symmetry potential in the following sections. 
   
What are the effects of the momentum-dependence of the symmetry potential on nucleon 
rapidity distributions? To answer this question we compare in the lower window of Fig.\ 7
nucleon rapidity distributions obtained with (MDI(0) and MDI(1)) and without ($MDYI+U_{sym}^{MDI(0)}$ 
and $MDYI+U_{sym}^{MDI(1)}$) the momentum-dependence in the symmetry 
potentials. It is seen that the momentum-dependence of the symmetry potential 
has very little effect on protons. This is mainly because the Coulomb potential 
works against and dominates over the symmetry potential on protons. 
However, the momentum-dependence of the symmetry potential 
affects significantly the neutron rapidity distributions. 
Overall, without the momentum-dependence of the symmetry 
potential (solid and dotted lines) the dN/Dy values for 
neutrons increases by 10 to 15\% with respect to the results with the momentum-dependent symmetry
potentials (long dashed and dashed lines). This increased emissions of neutrons are due to the 
stronger values of the symmetry potentials when the momentum-dependence is turned off as 
shown in Fig.\ 4. 

\subsection{Effects on isospin fractionation}
Isospin fractionation refers to the different isospin asymmetries in low and high density 
regions in nuclear reactions\cite{muller,liko,baran}. With symmetry energies increasing with density
it is energetically more favorable to have the low density regions more neutron-rich while the 
high density regions more isospin symmetric. The strength of the isospin fractionation 
has been found to depend sensitively on the density-dependence of the symmetry 
energy\cite{li00,liko,baran}. As we have discussed in the previous subsection, the symmetry
potential generally enhances the emission of neutrons while suppressing that of protons. 
A stiffer symmetry energy thus leads to a more neutron-rich low density gas of nucleons. 
Because of the charge and baryon number conservation, the remaining part (with local densities 
$\rho\geq\rho_0/8$) is consequently more isospin symmetric. It should be noted that
there is a net conversion of neutrons to protons because of the more abundant production of $\pi^-$ 
than $\pi^+$ above the pion production threshold. This effect, however, is very small 
even at a beam energy of 2 GeV/nucleon\cite{li03}. 
\begin{figure}[ht]
\centering \epsfig{file=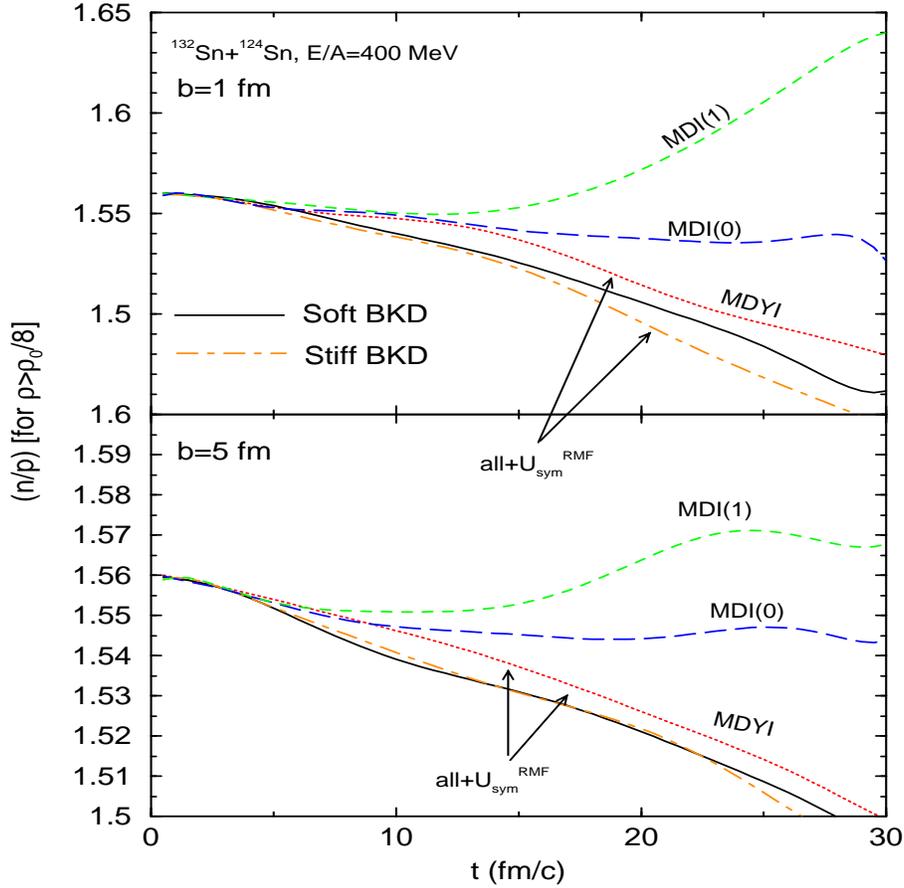,width=12cm,height=12cm,angle=-90} 
\vspace{1.cm}
\caption{{\protect\small Isospin asymmetry of bound nucleons in central (upper window) and
midcentral (lower window) collisions of $^{132}Sn+^{124}Sn$ at a beam energy of 
400 MeV/nucleon.}}
\label{figure8}
\end{figure}
To first explore effects of the symmetry potential on isospin fractionation
we study in Fig.\ 8 the isospin asymmetry (neutron/proton ratio) of bound nucleons as a 
function of time for central (upper window) and midcentral (lower window) collisions 
of $^{132}Sn+^{124}Sn$ at a beam energy of 400 MeV/nucleon. First of all, 
from the lowest three curves (the three calculations use three different isoscalar potentials
but the same symmetry potential of $U_{sym}^{RMF}$) in both windows, it is seen that the n/p ratio of 
bound nucleons (in the residue) is only slightly affected by the stiffness and the
momentum-dependence of the isoscalar potential. They are calculated with the
same symmetry potential $U_{sym}^{RMF}(\rho)$ but three different isoscalar potentials, the 
MDYI, soft BKD and the stiff BKD. There is only a variation of about 1-3\% in the 
$(n/p)_{\rho\geq \rho_0/8}$ ratio in the midcentral and central collisions. While it is very 
obvious that the $(n/p)_{\rho\geq \rho_0/8}$ ratio is much more sensitive to the variation of the 
symmetry potential. It is seen that the isospin asymmetry of bound nucleons increases 
by about 12\% and 5\% in central and midcentral collisions,  respectively, going from the 
stiffer to the softer symmetry energy (from the $U_{sym}^{RMF}$ through MDI(0) to the MDI(1)). 

What are the effects of the momentum-dependence of the symmetry potential on the isospin fractionation?
We study this question in Fig.\ 9 by comparing results obtained with and without 
the momentum-dependence of the symmetry potential. It is interesting to see that although 
the momentum-dependence of the symmetry potential increases the overall magnitude of isospin 
asymmetry of bound nucleons, relative effects of the symmetry energy remain about the same. 
In fact, in central collisions with the momentum-dependence of the symmetry potential
effects of the symmetry energy become even slightly stronger.    
\begin{figure}[ht]
\centering \epsfig{file=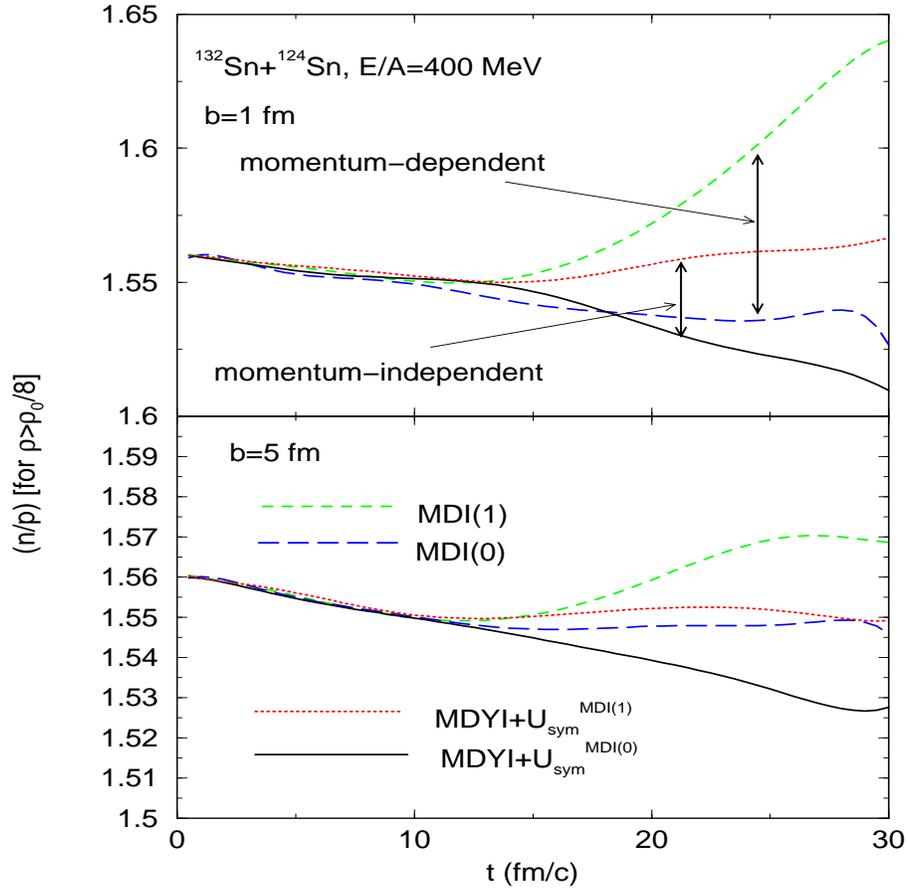,width=12cm,height=12cm,angle=-90} 
\vspace{1.cm}
\caption{{\protect\small Effects of the momentum-dependence of the symmetry potential on 
isospin asymmetry of bound nucleons in central (upper window) and
midcentral (lower window) collisions of $^{132}Sn+^{124}Sn$ at a beam energy of 
400 MeV/nucleon.}}
\label{figure8}
\end{figure}

\begin{figure}[ht]
\centering \epsfig{file=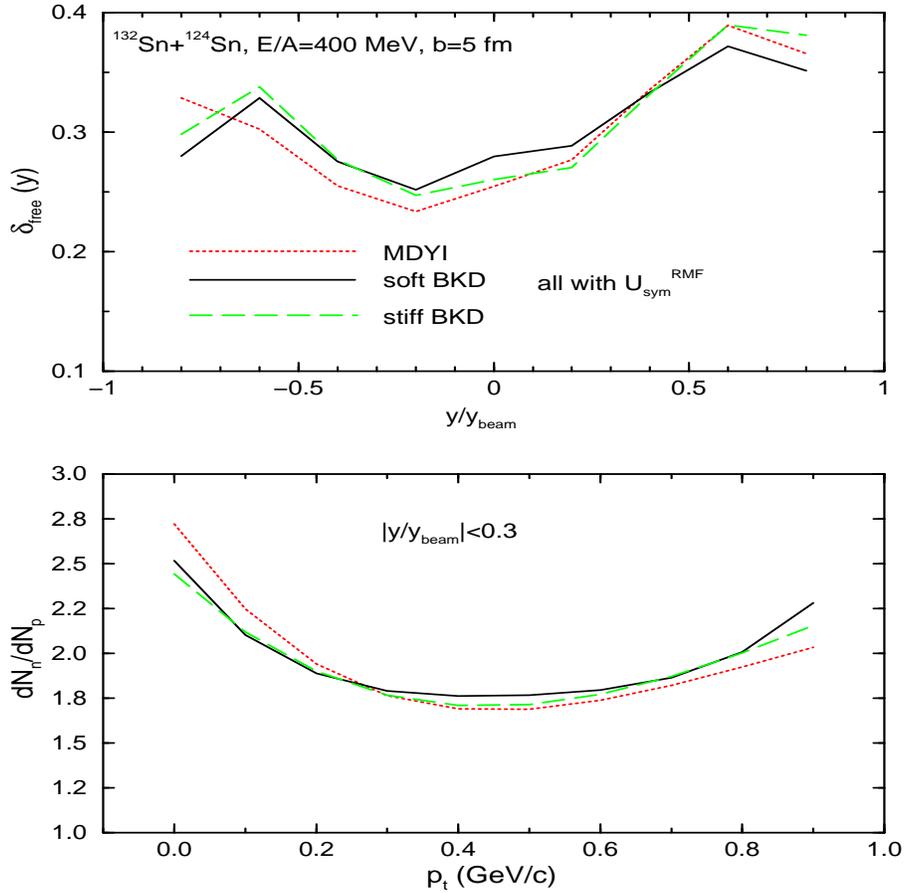,width=12cm,height=12cm,angle=-90} 
\vspace{1.cm}
\caption{{\protect\small Rapidity (upper window) and transverse momentum (lower window) 
distributions of isospin asymmetry of free nucleons with the three isoscalar 
potentials in the $^{132}Sn+^{124}Sn$ reaction at a beam energy of 
400 MeV/nucleon and an impact parameter of 5 fm.}}
\label{figure10}
\end{figure} 
\subsection{Combined constraints on density- and momentum-dependence 
of symmetry potentials}
Since the n/p ratio of bound or free nucleons are mainly affected by the symmetry potential unlike 
the inclusive observables which are strongly affected by isoscalar potentials, 
we concentrate in this subsection on the isospin asymmetry of free nucleons. Our main focus 
is to find observables that can be combined together to constrain 
simultaneously both the density and the momentum dependence of the symmetry potential. 
Specifically, we study the isospin asymmetry of free nucleons as a function of rapidity and 
transverse momentum. First, to show that these observables are indeed not so 
sensitive to the isoscalar potential we show in Fig.\ 10 the isospin asymmetry of 
free nucleons as a function of rapidity (upper window) and 
transverse momentum (lower window) calculated using the same symmetry potential $U_{sym}^{RMF}(\rho)$
but the three different isoscalar potentials MDYI, soft BKD and stiff BKD. It is seen that 
there is a variation of less than 8\% in the rapidity distribution of $\delta_{free}$
by changing from the MDYI to the soft BKD. The n/p ratio of free nucleons at midrapidity 
is shown as a function of transverse momentum in the lower window. For this observable 
the variation due to the change in isoscalar potential is even smaller at all transverse momenta. 
It should be mentioned that the overall rise of $dN_n/dN_p$ at low transverse momenta 
is due to the Coulomb repulsion on protons. 

How can we constrain both the density- and momentum-dependence of the symmetry potential? Shown in 
the upper window of Fig.\ 11 are the isospin asymmetries of free nucleons calculated with and 
without the momentum-dependence of the symmetry potential as a function of rapidity. First of all, 
there is a shift downward when the momentum-dependence of the symmetry potential is included for
both the soft (MDI(1)) and the stiffer (MDI(0)) symmetry energies. This is what we expected based on 
the comparisons of symmetry potentials in Fig.\ 4. Since the strength of the symmetry potential decreases
with the increasing momentum, the reduction of $\delta_{free}$ is particularly larger at higher rapidities.
It is also interesting to note that effects of the density-dependence of the symmetry potential, 
particularly stronger at midrapidity, are about the same with or without the momentum-dependence
of the symmetry potential. However, the value of $\delta_{free}$ with the soft symmetry 
energy but without the momentum-dependence of the symmetry potential (dotted line) 
is about the same as the one obtained using the momentum-dependent stiff symmetry energy 
MDI(0) (long dashed) at midrapidity, although they are quite different at forward and backward rapidities.
\begin{figure}[ht]
\centering \epsfig{file=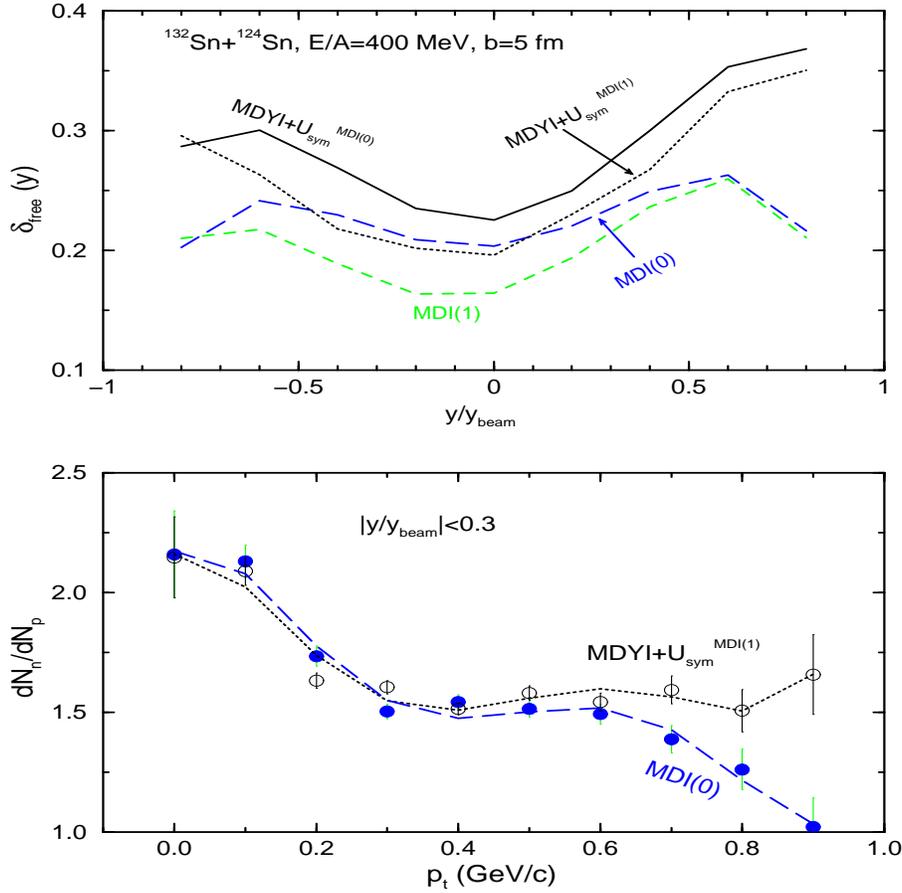,width=12cm,height=12cm,angle=-90} 
\vspace{1.cm}
\caption{{\protect\small Rapidity and transverse momentum distributions of 
isospin asymmetry of free nucleons with and without the momentum dependence of the 
symmetry potential in the $^{132}Sn+^{124}Sn$ reaction at a beam energy of 
400 MeV/nucleon and an impact parameter of 5 fm.}}
\label{figure10}
\end{figure}
How do we separate them at midrapidities? One possible way is shown in the 
lower window of Fig.\ 11 where we compare 
the ratio of free neutrons to protons at midrapidity as a function of transverse momentum with the MDI(0) 
and the MDYI+$U_{sym}^{MDI(1)}(\rho)$ interactions. It is seen that the two interactions 
are clearly separated at high transverse momenta.   
This is mainly because the symmetry potential $U_{sym}^{MDI(1)}(\rho)$ remains a constant while that 
embedded in the MDI(0) interaction decreases with increasing momenta as shown in Fig.\ 4.
Thus by combining the measurements of the rapidity and transverse momentum distributions of the isospin 
asymmetry of free nucleons one can determine simultaneously the density- and momentum-dependence of the 
symmetry potential. The density dependence of the symmetry energy can then be determined uniquely. 
We stress here that the separation at high transverse momenta is not due to the weaker isoscalar potential
at high $p_T$ with the MDI(0) interaction as shown in Fig. 5. This is because the isoscalar potential acts
equally on both neutrons and protons, the ratio $dN_n/dN_p$ is thus insensitive to the EOS of 
symmetric nuclear matter as shown in Fig. 10. 
\begin{figure}[ht]
\centering \epsfig{file=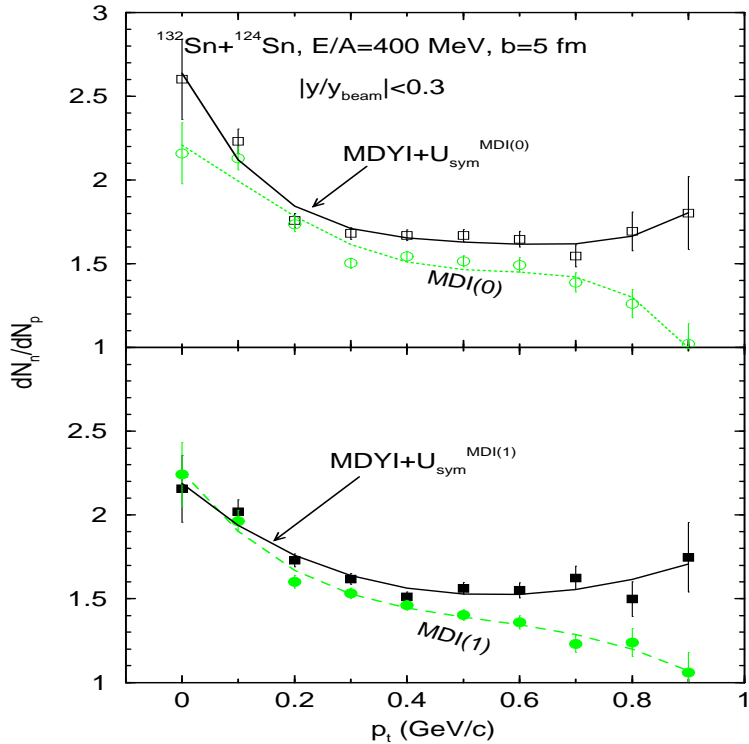,width=10cm,height=10cm,angle=-90} 
\vspace{1.cm}
\caption{{\protect\small Transverse momentum distributions of 
isospin asymmetry of free nucleons with and without the momentum dependence of the 
stiff (upper window) and soft (lower window) symmetry potential in the 
$^{132}Sn+^{124}Sn$ reaction at a beam energy of 
400 MeV/nucleon and an impact parameter of 5 fm.}}
\label{figure11}
\end{figure} 
For a given density-dependent symmetry energy $E_{sym}(\rho)$, how does the momentum-dependence of the 
symmetry potential affect the $p_t$ distribution of $dN_n/dN_p$? To answer this question we 
compare in Fig.\ 12 the $dN_n/dN_p$ distributions with the stiff (upper window) and soft (lower window)
symmetry energy with (open and filled circles) and without (open and filled squares) the momentum-dependence
of the symmetry potential. It is seen that without the momentum-dependence of the symmetry potential, 
the $dN_n/dN_p$ ratio turns to saturate above about $p_t\geq 0.3 GeV/c$. With the momentum-dependence of
the symmetry potential, however, the $dN_n/dN_p$ continue to decrease to high $p_t$ values. Again, this 
is a direct reflection of the decreasing symmetry potential with the increasing momentum. 
From both Fig.\ 11 and Fig.\ 12 one sees clearly the importance of measuring the isospin 
asymmetry of high $p_t$ particles at midrapidity.

\subsection{Effects on rapidity and transverse momentum dependence of free nucleons in central collisions}
In the previous subsection, we concentrated on midcentral collisions at an impact parameter of 5 fm.
To examine the impact parameter dependence, we examine effects of the momentum-dependence of 
the symmetry potential in the $^{132}Sn+^{124}Sn$ reactions at 400 MeV/nucleon and 
an impact parameter of 1 fm with the soft and stiff symmetry energy in Fig.\ 13 
and Fig.\ 14, respectively.
Comparing the results at an impact parameter of b=1 fm in Fig. 13 and Fig. 14 with those at an impact 
parameter of b=5 fm in the previous subsection, it is seen that the main effects of the 
momentum-dependence of the symmetry potential do not change that much. This is mainly because 
for such a heavy reaction system as the $^{132}Sn+^{124}Sn$, the variation of impact parameter 
from 1 fm to 5 fm does not result in a very appreciable change in both the density and 
the isospin asymmetry in the participant region during the reaction. More quantitatively, 
with the soft symmetry energy MDI(1) the effects of the momentum-dependence of the 
symmetry potential at b=1 fm are somewhat stronger than that at b=5 fm. However, with the
stiff symmetry energy the impact parameter dependence is less obvious. This is understandable from the 
strength of the symmetry potentials shown in Fig. 4. It is seen there that the slope of 
the symmetry potential depends more strongly on the density in the case of the soft 
symmetry energy (MDI(1)). Thus, as the density increases due to the increasing centrality,
the force due to the symmetry potential increases more strongly with the soft symmetry energy, 
leading to the larger effects shown in Fig.\ 13. 
\begin{figure}[ht]
\centering \epsfig{file=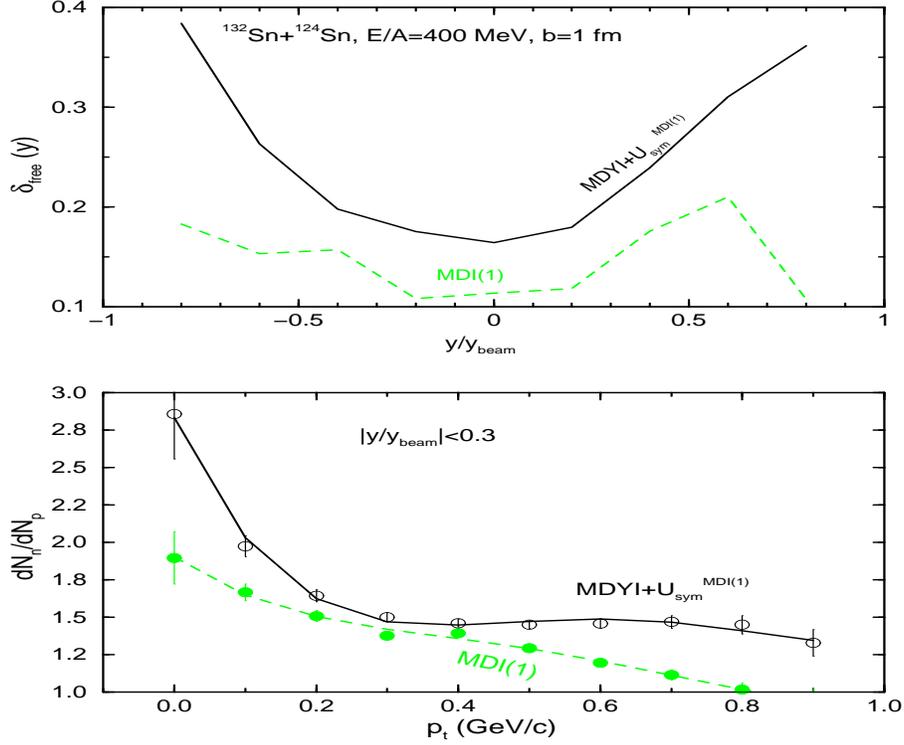,width=10cm,height=12cm,angle=-90} 
\vspace{1.cm}
\caption{{\protect\small Rapidity (upper window) and transverse momentum (lower window) 
distributions of isospin asymmetry of free nucleons with and without the momentum 
dependence of the {\it soft} symmetry energy in the $^{132}Sn+^{124}Sn$ reaction at a 
beam energy of 400 MeV/nucleon and an impact parameter of 1 fm.}}
\label{figure12}
\end{figure}
\begin{figure}[ht]
\centering \epsfig{file=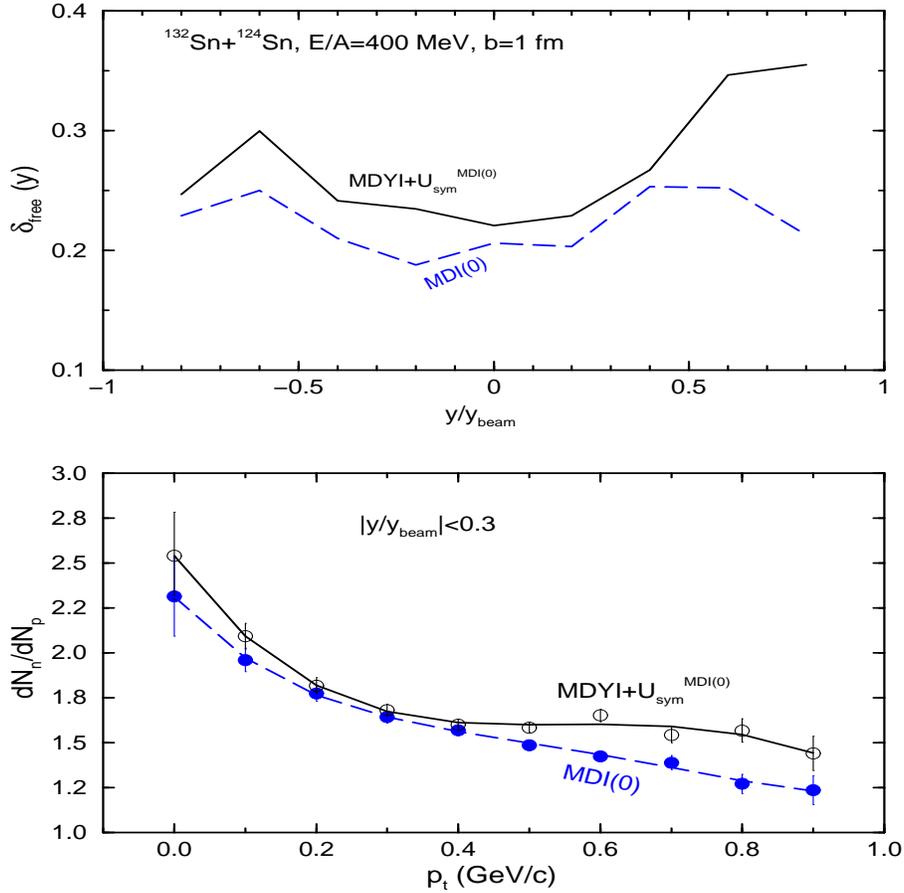,width=12cm,height=12cm,angle=-90} 
\vspace{1.cm}
\caption{{\protect\small Rapidity (upper window) and transverse momentum (lower window) 
distributions of isospin asymmetry of free nucleons with and without the momentum 
dependence of the {\it stiff} symmetry energy in the $^{132}Sn+^{124}Sn$ reaction at a 
beam energy of 400 MeV/nucleon and an impact parameter of 1 fm.}}
\label{figure13}
\end{figure}
\subsection{Effects on proton direct flow and neutron-proton differential flow}
How does the momentum-dependence of the symmetry potential affect the nuclear collective flow?
First of all, it is important to know what collective observables are sensitive to the 
symmetry potential with or without the momentum-dependence. Since symmetry potential is 
small compared to the isoscalar potential, moreover, at high energies baryon-baryon collisions 
dominate the reaction dynamics, one has to use delicate observables to extract information about the
symmetry potential. To illustrate this point, we show in Fig. 15 the standard transverse flow
analysis for protons in the reaction of $^{132}Sn+^{124}Sn$ at a beam energy of 400 MeV/nucleon and an
impact parameter of 5 fm. It is seen that with (upper window) or without (lower window) the
momentum-dependence of the symmetry potential for both the soft and stiff symmetry energies, 
the proton transverse flow is about the same. 
\begin{figure}[ht]
\centering \epsfig{file=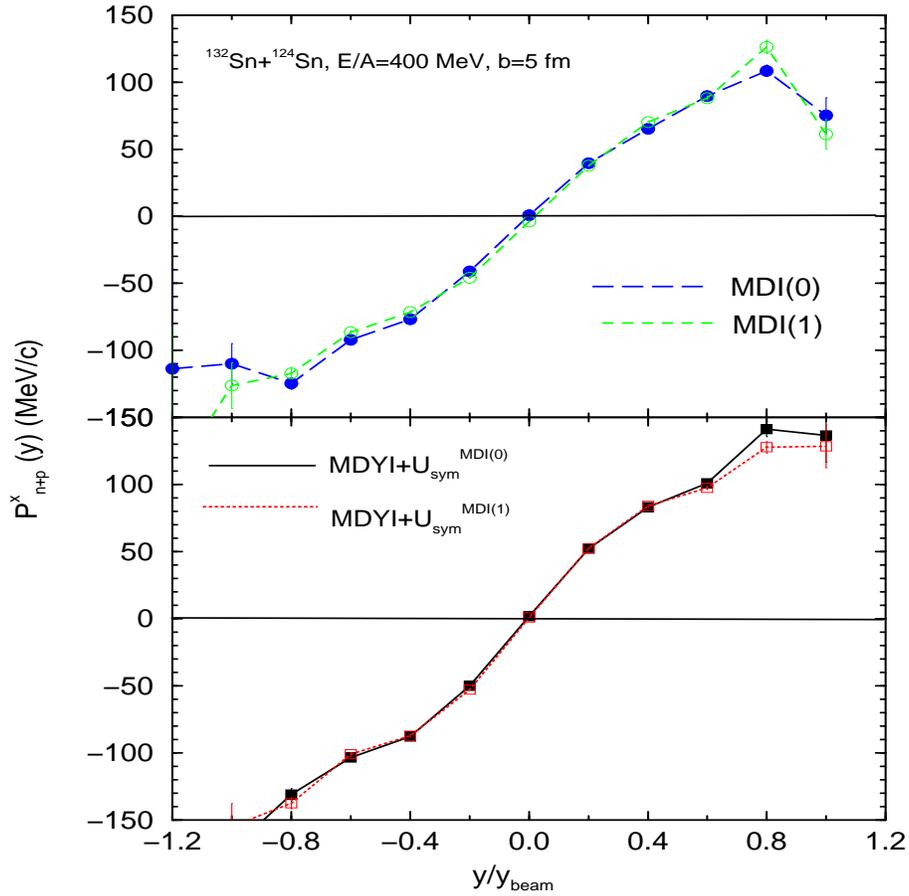,width=12cm,height=12cm,angle=-90} 
\vspace{1.cm}
\caption{{\protect\small Proton transverse flow analysis with (upper window) and without (lower window)
the momentum-dependence of the symmetry potential  
in the reaction of $^{132}Sn+^{124}Sn$ at a beam energy of 400 MeV/nucleon 
and an impact parameter of 5 fm.}}
\label{figure14}
\end{figure}

\begin{figure}[ht]
\centering \epsfig{file=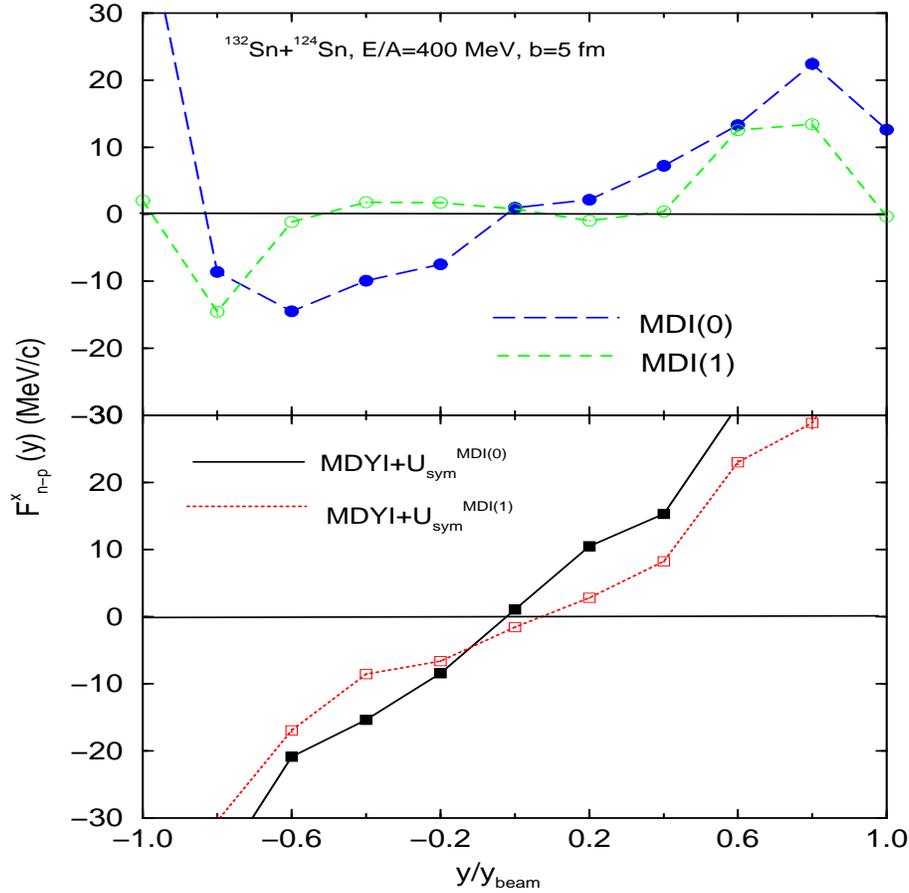,width=12cm,height=12cm,angle=-90} 
\vspace{1.cm}
\caption{{\protect\small Neutron-proton differential flow analysis 
with (upper window) and without (lower window)
the momentum-dependence of the symmetry potential  
in the reaction of $^{132}Sn+^{124}Sn$ at a beam energy of 400 MeV/nucleon 
and an impact parameter of 5 fm.}}
\label{figure15}
\end{figure}
We now turn to the neutron-proton differential flow 
\begin{equation}
F^x_{n-p}(y)\equiv\sum_{i=1}^{N(y)}(p^x_iw_i)/N(y),
\end{equation}
where $w_i=1 (-1)$ for neutrons (protons) and $N(y)$ is the total number of free 
nucleons at rapidity $y$. The differential flow combines constructively effects of the 
symmetry potential on the isospin fractionation and the collective flow. 
It has the advantage of maximizing effects of the symmetry potential while minimizing effects
of the isoscalar potential\cite{li00}. Shown in Fig.\ 16 is the neutron-proton differential flow
with (upper window) and without (lower window) the momentum-dependence of the symmetry potential. 
First, it is seen that in both cases, the neutron-proton differential flow is rather sensitive
to the symmetry energy compared to the standard proton transverse flow analysis shown in Fig. 15.
Secondly, it is seen that the neutron-proton differential flow is stronger without the momentum-dependence
of the symmetry potential. This is because the momentum-independent symmetry potential 
makes not only more neutrons free but also gives them higher transverse momenta in the 
reaction plane because of its higher magnitude and steeper 
slopes compared to the momentum-dependent symmetry potential embedded in eq.\ \ref{mdi}.

Is the neutron-proton differential flow still sensitive to the variation of $E_{sym}(\rho)$ 
when the momentum-dependent symmetry potential is used? The answer is a definitely yes
as shown in the upper window of Fig.\ 16. In fact, the slope of the differential 
flow $F^x_{n-p}(y)$ at y=0 changes 
from about -1.9 to 21.4 MeV/c using eq. \ref{mdi} by changing 
from the MDI(1) to the MDI(0) parameter set, while it changes from about 
26 to 39 MeV/c with the momentum-independent symmetry potentials. 
The net change due to the variation of the $E_{sym}(\rho)$ 
is thus actually larger with the momentum-dependent symmetry potentials. 
Thus the neutron-proton differential flow remains a sensitive probe to the density 
dependence of the symmetry energy.

\section{summary}  
In summary, using an isospin- and momentum-dependent transport model we studied 
effects of the momentum-dependent symmetry potential on several experimental observables
in heavy-ion collisions induced by neutron-rich nuclei at RIA energies. 
The momentum-dependence of the symmetry potential is found to play an important role
in the isospin fractionation, particle emission and neutron-proton differential flow.
In particular, high $p_t$ neutron to proton ratio at midrapidity is very sensitive to the 
momentum-dependence of the symmetry potential. 
However, the momentum-dependence of the symmetry potential has little effects on the overall 
reaction dynamics and the directed nuclear flow. To uniquely determine the 
$E_{sym}(\rho)$ one needs to determine simultaneously both the density- and momentum-dependence 
of the symmetry potential. For this purpose, it is necessary to
combine several experimental observables, such as, the rapidity and transverse momentum 
dependence of the free neutron to proton ratio, and the neutron-proton differential flow.  
Comparing experimental data with predictions based on transport models including the 
momentum-dependent symmetry potential is thus crucial for investigating accurately 
the ${\rm EOS}$ of dense neutron-rich matter. 

\section{acknowledgement}
This research is supported in part by the National Science Foundation of the 
United States under grant No. PHY-0088934, PHY-0243571, the Natural Sciences and 
Engineering Research Council of Canada, and the Fonds Nature et Technologies of Quebec.

\end{document}